\documentclass[12pt,twoside,a4paper,fleqn]{article}
\usepackage{amsmath}
\usepackage{epsfig}
\usepackage{subfigure}
\usepackage{here}
\usepackage{float}
\usepackage{bm}
\usepackage[font=sf,labelfont=bf,parskip=3pt,width=.97\textwidth]{caption}
\usepackage{subfigure}

\begin{document}

\title{ 
\vskip-3cm 
{\normalsize
\hfill MZ-TH/07-22 \\
\hfill \today \\
}
\vskip4cm 
{\bf Determination of QCD condensates}\\
{\bf from $\tau$-decay data}
\\[3ex] 
}

\date{}
\author{ 
A.~A.~Almasy\footnote{Email: andrea.almasy@desy.de} \\[1ex]
\normalsize{II. Institut f\"ur Theoretische Physik, 
 Universit\"at Hamburg}\\ 
\normalsize{Luruper Chaussee 149, D-22761 Hamburg, Germany} \\[5ex]
K.~Schilcher,~ 
H.~Spiesberger\footnote{Email: hspiesb@thep.physik.uni-mainz.de} \\[1ex]
\normalsize{Institut f\"ur Physik, Johannes-Gutenberg-Universit\"at,}\\ 
\normalsize{Staudinger Weg 7, D-55099 Mainz, Germany} \\[10ex]
}

\maketitle

\begin{abstract}
\medskip 
\noindent 
We have used the latest data from the ALEPH Collaboration to extract
 values for QCD condensates up to dimension $d=12$ in the $V-A$ channel
 and up to dimension $d=8$ in the $V$, $A$ and $V+A$ channels. 
 Performing 2- and 3-parameter fits, we obtain new results for the 
 correlations of condensates. The results are consistent among 
 themselves and agree with most of the previous results found in the 
 literature. 
\end{abstract}

\thispagestyle{empty}

\clearpage


\section{Introduction} 

QCD is widely considered to be a good candidate for a theory of the
 strong interactions. Asymptotic freedom allows us to perform a
 perturbative treatment of strong interactions at short distances.  Long
 distance behaviour, however, is not fully understood: it is commonly
 believed that, due to the nontrivial structure of the physical vacuum,
 the perturbation expansion does not completely define the theory and
 that one has to add non-perturbative effects as well. In order to make
 a comparison with experiments possible even in the resonance energy
 range, Shifman, Vainshtein and Zakharov \cite{shifman1} have proposed
 to use the Operator Product Expansion (OPE) and to introduce the vacuum
 expectation values of operators occurring in the OPE, the so called
 {\em condensates}, as phenomenological parameters. It is worth to study
 these parameters in order to see whether one can indeed obtain a
 consistent description of the low energy hadronic physics and get more
 insight into the properties of the QCD vacuum. It is, in particular,
 important to determine the range of values of the condensates allowed
 by presently available experimental data.
 
Condensates are needed in the description of two-point functions of
 hadronic currents: together with the results of perturbative QCD they
 allow us to obtain approximate theoretical predictions for the hadronic
 current correlators in the space-like region. On the other hand, in the
 time-like region, the discontinuity of these amplitudes is directly
 related to measurable quantities. Analyticity strongly correlates the
 energy dependence of the amplitudes in these two regions; however, due
 to errors affecting both the theoretical predictions as well as the
 experimental data, the relation of the two-point functions in the
 time-like and space-like regions must be carefully analysed before it
 can be used in applications.
 
There are several methods, generically called {\em QCD sum rules}
 \cite{shifman1,bell,bertlmann}, that have been used in the past for
 obtaining values of the QCD condensates. They all rely implicitly on
 the assumption that an analytic extrapolation from the time-like to the
 space-like region is possible without introducing additional
 uncertainties. Therefore they can include {\em theoretical errors} in
 the space-like region only at a qualitative level, and/or need
 (explicit and implicit) assumptions on the derivatives of the
 amplitudes. A quantitative estimate of the errors including both
 experimental and theoretical ones (truncation of the perturbative and
 operator product expansions) is therefore very difficult in these
 approaches. The application of fully controlled analytic extrapolation
 techniques should remedy these effects. There are a few methods of this
 sort, in which the error channels in the space-like region are defined
 through $L^2$-norms \cite{almasy,ciulli} or $L^\infty$-norms
 \cite{auberson,auberson2,causse}.
 
The functional method that we have developed and used in a previous
 publication for an analysis of the correlator of the $V-A$ current
 \cite{almasy}, allows us to extract within rather general assumptions
 the condensates from a comparison of the time-like experimental data
 with the asymptotic space-like results from theory. We will see that
 the price to be paid for the generality of assumptions is relatively
 large errors in the values of the extracted parameters. In this
 respect, our method is not superior to other approaches; however, we
 hope that our results provide additional confidence in the numerical
 results obtained with the help of methods based on QCD sum rules
 \cite{shifman1,bell,bertlmann,bell1,bertlmann1,dominguez0}.
 
The paper is organised as follows: In the first section the theoretical
 aspects of hadronic $\tau$-decays are considered while in section 3 we
 describe the experimental data at hand. The actual method used to
 extract the condensates is presented in section 4. Basically the method
 is the one reported in \cite{almasy} but here a generalisation to all
 channels is performed. The results are summarised in section 5. We
 quote results in all the four channels where data are available.
 Results for the $V-A$ channel were already published in \cite{almasy};
 for completeness, we repeat those previous results in the present paper
 and extend our analysis by a new 3-parameter fit. New results were
 obtained in the $V$, $A$ and $V+A$ channels from 1-, 2- and 3-parameter
 fits. We shall also discuss important consistency checks of our method 
 in section 6. The comparison of our findings with others present in the 
 literature is performed in section 7.


\section{Theoretical description of hadronic $\tau$-decays}
\label{sec:theory}

The $\tau$ lepton is heavy enough ($m_\tau = 1.777$ GeV) to decay not 
 only into other leptons, but into final states involving hadrons as well. 
 These decays offer an ideal laboratory for the study of strong 
 interactions, including the transition from the perturbative to the 
 non-perturbative regime of QCD in the simplest possible reaction. This 
 might explain the tremendous efforts ongoing in both theoretical and 
 experimental studies of $\tau$ physics.

We consider the correlator of hadronic vector and axial-vector charged
 currents, $J_{\mu} = V_{\mu} = \bar{u}\gamma _{\mu}d$ and $J_{\mu} =
 A_{\mu} = \bar{u} \gamma_{\mu}\gamma_{5}d$,
\begin{eqnarray}
 \Pi _{\mu \nu }^{J} 
 & = & 
 i\int d^4xe^{iqx}\langle TJ_{\mu }(x)J_{\nu}(0)^{\dagger}\rangle 
 \\
 & = & 
 \left( -g_{\mu \nu } q^{2} + q_{\mu }q_{\nu } \right) 
 \Pi_{J}^{(0+1)}(q^{2}) + g_{\mu \nu } q^2 \Pi _{J}^{(0)}(q^{2}) \,.
\nonumber
\end{eqnarray}
The conservation of the vector current implies $\Pi_{V}^{(0)}=0$. The
connection to experimental observables is most easily expressed with the
help of the spectral functions which are related to the absorptive part
of the correlators. Using the normalisation as defined in most of the
previous publications, the functions
\begin{equation}
v_{j}(s) = 2\pi {\rm Im}\Pi_{V}^{(j)}(s),~~~~a_{j}(s)
         = 2\pi {\rm Im}\Pi_{A}^{(j)}(s) 
\label{specfun}
\end{equation}
can be extracted from the decay spectrum of hadronic $\tau $-decays.

The hadronic polarisation tensor can be rewritten using the OPE:
\begin{equation}
 \Pi^{(j)}_{V,A}(s) =
 \sum_{d\ge0}\frac{{\cal O}_d^{(j),{V,A}}}{(-s)^{d/2}},
\label{opeexp}
\end{equation}
where ${\cal O}_d\equiv {\cal C}_d\langle{\cal O}_d\rangle$ is the short
 hand notation for the QCD non-perturbative condensate 
$\langle{\cal O}_d\rangle$ of dimension $d$ and its associated
 perturbative Wilson coefficient ${\cal C}_d$; $s\equiv q^2$ is
 the squared momentum transfer.

The contribution to (\ref{opeexp}) of lowest dimension, $d=0$, is 
 entirely given by perturbation theory. For that reason it is useful 
 to separate the two contributions in (\ref{opeexp}):
\begin{equation}  
 \Pi^{(j)}_{V,A}(s) =
 \Pi_{{\rm PT},V,A}^{(j)}(s) + \Pi_{{\rm OPE},V,A}^{(j)}(s).
\label{corrseparation}
\end{equation}

As will become clear when we describe our method (section \ref{sec:method})
 it is enough to consider the first derivative of the perturbative part, 
 i.e., the Adler function
\begin{equation}
 D(s) \equiv -s \frac{d}{ds}\Pi_{\rm PT}(s),
\end{equation}
which is known in the massless-quark limit up to terms of order
 $\alpha_s^4$. After re-summing the leading logarithms it reads, for
 space-like momenta ($s<0$): 
\begin{equation}
 D_{V,A}(s) = \frac{1}{4\pi^2}
 \sum_{n\ge0}^4\ K_n \left(\frac{\alpha_s(-s)}{\pi}\right)^n .
\end{equation}
The coefficients $K_n$ are the same for both $V$ and $A$ channels. For 3
 flavours, in $\overline{\rm MS}$ regularisation, $K_0=K_1=1$, $K_2=1.64$ 
 \cite{chetyrkin2,celmaster,dine}, $K_3 = 6.37$ \cite{gorishny,surguladze} 
 and for $K_4$ there are two estimates $K_4 = 25 \pm 25$ \cite{kataev} and 
 $K_4=27\pm16$ \cite{baikov}.

For the correlators with spin $0+1$ and in the chiral limit, the
 non-perturbative part has the form
\begin{equation}
 \Pi^{(0+1)}_{{\rm OPE},V,A}(s) =
 \sum_{d\ge4}\frac{{\cal O}_d^{V,A}}{(-s)^{d/2}}
 \left(1 + c_d^{{\rm NLO}, V,A} \frac{\alpha_s(\mu^2)}{\pi} \right),
\label{OPE:0+1}
\end{equation}
where perturbative corrections of order $O(\alpha_s)$ are taken into 
 account, described by coefficients $c_d^{\rm NLO}$. Some of the NLO 
 coefficients were calculated in \cite{chetyrkin,adam,lanin}. 

The parameters ${\cal O}_d$ can be related to vacuum expectation values
 of products of quark and gluon field operators \cite{shifman1}. Often
 vacuum dominance or the factorization approximation, which holds, e.g.,
 in the large-$N_c$ limit, is assumed. Our analysis does not rely on
 such a representation.


\section{Experimental data}

Since its discovery, the $\tau$ lepton has been studied with
 ever-increasing precision at every new $e^+e^-$ collider that has gone
 into operation. We are particularly interested in the comprehensive 
 measurements of exclusive hadronic branching ratios from ALEPH 
 \cite{aleph,aleph05} and of the non-strange spectral functions from 
 ALEPH \cite{aleph,aleph05} and OPAL \cite{opal} that have yielded 
 important contributions to studies of perturbative QCD at low energies 
 and, in particular, to the measurement of $\alpha_s(m_\tau^2)$. Recent 
 measurements of a set of semi-exclusive branching ratios by DELPHI are 
 also available \cite{delphi}, but have not been analyzed to a similar 
 extent. A number of exclusive branching ratio measurements from BaBar 
 have been reported already \cite{babar} and with more work invested in 
 the understanding of these higher-multiplicity final states we may 
 expect high-precision data for the spectral functions to come also from 
 the B-factories.

We have chosen to use the final data from the ALEPH collaboration
 \cite{aleph05} because, as compared to those available from OPAL
 \cite{opal}, they have the smallest experimental errors and provide 
 a larger number of bins. 

The spectral functions (\ref{specfun}) are obtained by dividing the 
 normalised invariant mass-squared distribution of hadronic $\tau$ 
 decays, $dR_{\tau,{V,A}}/ds$, for a given hadronic mass $\sqrt s$ by 
 the appropriate kinematic factor. They are then normalised to the 
 branching fraction of the massless leptonic, i.e.\ electron, channel 
 ${\cal B}_e=(17.810\pm0.039)\%$ \cite{aleph05}:  
\begin{equation}
\begin{array}{r}
\displaystyle v_1(s)=\frac{m_\tau^2}{6|V_{ud}|^2S_{\rm EW}}
\frac{dR_{\tau,{V}}}{{\cal B}_eds}\left[\left(1-\frac{2}{m_\tau^2}\right)^2
\left(1+2\frac{s}{m_\tau^2}\right)\right]^{-1},\\
\\
\displaystyle a_1(s)=\frac{m_\tau^2}{6|V_{ud}|^2S_{\rm EW}}
\frac{dR_{\tau,{A}}}{{\cal B}_eds}\left[\left(1-\frac{2}{m_\tau^2}\right)^2
\left(1+2\frac{s}{m_\tau^2}\right)\right]^{-1},\\
\\
\displaystyle a_0(s)=\frac{m_\tau^2}{6|V_{ud}|^2S_{\rm EW}}
\frac{dR_{\tau,{A}}}{{\cal B}_eds}\left(1-\frac{2}{m_\tau^2}\right)^{-2}.
\end{array}
\label{specdef}
\end{equation}
Here $S_{\rm EW}=1.0198\pm0.0006$ accounts for short distance
 electroweak radiative corrections \cite{marciano} and the CKM mixing
 matrix element has the value $|V_{ud}|=0.9746\pm0.0006$ \cite{davier}.
 Due to the conservation of the vector current, there is no $j=0$
 contribution to the vector spectral function, while the only
 contribution to $a_0$ is assumed to come from the pion pole. The 
 spectral function $a_0$ is connected, via partial conservation of the 
 axial-vector current (PCAC), to the pion decay constant 
 $f_\pi=0.1307\mbox{ GeV}$ \cite{partdatagroup} through 
 $a_{0,\pi}(s)=2\pi^2f_\pi^2\delta(s-m_\pi^2)$.


\section{The determination of condensates: a functional method}
\label{sec:method}

Let us consider a set of functions $F(s)$ which are admissible as a
 representation of the true correlator if they are real analytic
 functions in the complex $s$-plane cut along the time-like interval
 $[s_0,\infty)$ with $s_0>0$. The asymptotic behaviour of $F(s)$ is 
 restricted by fixing the number of subtractions in the dispersion 
 relation between $F(s)$ and its imaginary part $f(s)=\mbox{Im} F(s+i0)$ 
 along the cut: 
\begin{equation}
 F(s)=\frac{1}{\pi}\int_{s_0}^\infty \frac{f(z)dz}{z-s}+
 \mbox{subtractions}\ .
\label{disprel}
\end{equation}

For our purpose it is convenient to get rid of the subtraction terms by
 taking an appropriate number of derivatives with respect $s$. We denote 
 by ${\cal C}_n(s,z)$ the kernel occurring in the dispersion relation for 
 the $n$-th derivative of $F(s)$: 
\begin{equation}
 -s^nF^{(n)}(s) =
 \frac{1}{\pi}\int_{s_0}^\infty {\cal C}_n(s,z)f(z)dz\ .
\label{disprel-n}
\end{equation}
For example, in the special case of the $V-A$ correlator which
 vanishes identically in the chiral limit to all orders in QCD
 perturbation theory there are no subtractions needed and thus one takes
 $n=0$. In this case, the dispersion relation will be identical to the one
 of Eq.\ (\ref{disprel}) with no subtraction terms. On the other hand, the
 $V$, $A$ and $V+A$ correlators are dominated by their perturbative
 contributions and there is one subtraction needed in the dispersion
 relation. To get rid of this usually unknown constant term one needs to
 take the first derivative and set $n=1$.

In order to determine $F(s)$ and $f(s)$ we use the following two
 available sources of information:
\begin{itemize}
\item experimental data measured in the time-like interval 
 $\Gamma_{\rm exp}=[s_0,s_{\rm max}]$: 
 \begin{equation}
 f_{\rm exp}(s)=\frac{1}{2\pi}\left\{
 \begin{array}{ll}
 v_1(s), & \mbox{in the}\ V\ \mbox{channel,}\\
 a_1(s)+a_0(s), & \mbox{in the}\ A\ \mbox{channel.}
 \end{array}
 \right. 
 \end{equation}
 The extension to $V\pm A$ channels is straightforward.
\item theoretical model given by perturbative QCD, i.e.,
 \begin{itemize}
 \item the prediction for $F(s)$ in the space-like interval 
 $\Gamma_L = [s_2,s_1]$: 
 \begin{equation}
 F_{\rm QCD}(s) \equiv \Pi^{(0+1)}_J(s), 
 \quad J = V,~A,~V+A,~V-A
 \end{equation}
\item and $\left. f_{\rm QCD}(s) = \mbox{Im}F_{\rm QCD}(s+i0)
 \right|_{s\in(s_{\rm max},\infty)}$ since QCD is expected to be reliable 
 for large energies. 
\end{itemize}
\end{itemize}

As a next step in extracting values for the condensates, we split the
 integral on the r.h.s.\ of the modified dispersion relation
 (\ref{disprel-n}) into two parts: one that can be described by the
 experiment and the other one by the theoretical model, i.e., QCD:
\begin{equation}
\underbrace{-s^nF^{(n)}_{\rm QCD}(s) - \frac{1}{\pi}
\int_{s_{\rm max}}^\infty {\cal C}_n(s,z)f_{\rm QCD}(z)dz} =
\underbrace{\frac{1}{\pi}\int_{s_0}^{s_{\rm max}}{\cal C}_n(s,z)f(z)dz}.
\label{dispsource}
\end{equation}
\vspace{0.1 cm}
\hspace{2.0 cm}{QCD prediction: ${\tilde F}^n_{\rm QCD}(s)$ }
\hspace{2.3 cm}{experimental data}
\vspace{0.5 cm}
 
The goal of the method is to check if there exists a function $F(s)$
 which is in accord with both the data on $\Gamma_{\rm exp}$ and the
 model on $\Gamma_L$. For doing this, one can use an $L^2$-norm approach
 and define two functionals $\chi_L^2[f]$ and $\chi_R^2[f]$.
 $\chi_R^2[f]$ compares the true amplitude $f(s)$ with the data. Here
 one can take into account not only experimental errors on each
 individual bin, but the full correlation of available data by using the
 covariance matrix $V$ provided by ALEPH as a weight function. Therefore
 we define
\begin{equation}
 \chi^2_R[f] = \frac{1}{|\Gamma_{\rm exp}|}\int_{s_0}^{s_{\rm max}}dx
 \int_{s_0}^{s_{\rm max}}dx'
 V^{-1}(x,x')(f(x)-f_{\rm exp}(x))(f(x')-f_{\rm exp}(x')).
\end{equation}

As a measure for the agreement of the true function $f(s)$ with the
 theory, we define $\chi_L^2[f]$ by comparing the left and right hand
 sides of (\ref{dispsource})
\begin{equation}
\chi^2_L[f] = \displaystyle
\frac{1}{|\Gamma_L|}\int_{\Gamma_L}w_L(x)
\left({\tilde F}^n_{\rm QCD}(x) -
\displaystyle \frac{1}{\pi}
\int_{s_0}^{s_{\rm max}}{\cal C}_n(x,x')f(x')dx'\right)^2dx,
\label{chiL2}
\end{equation}
where $w_L$ is a weight function for the space-like interval, i.e.\ an
 {\em a-priori} estimate of the accuracy of the QCD predictions, and
 written as $1/\sigma_L^2(s)$. $\sigma_L(s)$ should be chosen as a 
 continuous, strictly positive function of $s\in\Gamma_L$ and encodes 
 errors due to the truncation of the perturbative series and the OPE. It 
 is expected to decrease as $|s|\rightarrow\infty$ and diverge for
 $s\rightarrow0$. For example, in the case of the $V-A$ correlator
 we will use the next higher dimension contribution in the OPE as an
 error estimate on the space-like region. In the case of the $V$,
 $A$ and $V+A$ correlators the situation is a bit more complicated since 
 they are dominated by their perturbative part. Thus one has three 
 possibilities to define an error corridor: one can use the last known 
 term of the perturbation series, or the first omitted term in the OPE, 
 or a combination of the two of them. As an illustration, in the case of 
 a 1-parameter fit $\sigma_L^{V,A}(s)$ would be given by
\begin{equation}
 \sigma_L^{V,A}(x) = \left\{
\begin{array}{ll}
 \displaystyle\frac{1}{4\pi^2}K_3\left(
\frac{\alpha_s(-x)}{\pi}\right)^3,\\
\\
\displaystyle3\frac{{\cal O}_6^{V,A}}{(-x)^3} \, , \\
\\
 \displaystyle \sqrt{\left[\frac{1}{4\pi^2}K_3
 \left(\frac{\alpha_s(-x)}{\pi}\right)^3\right]^2
 + \left[3\frac{{\cal O}_6^{V,A}}{(-x)^3}\right]^2} \, . \\
\end{array}
\right.
\label{errcorr}
\end{equation}
The factor 3 in front of ${\cal O}_6^{V,A}$ in the expression of 
 $\sigma_L^{V,A}(x)$ comes from the fact that in the $V$, $A$ and $V+A$ 
 channels one needs to set $n=1$ in Eq.\ (\ref{disprel-n}) and thus take 
 the first derivative of the operator product expansion in Eq.\ 
 (\ref{OPE:0+1}). $K_3$ was given in section \ref{sec:theory}.

In order to find the true function $f(s)$ one can combine the
 information contained in these two functionals by means of Lagrange
 multipliers and find the unrestricted minimum of
\begin{equation}
{\cal F}[f] = \chi^2_L[f] + \mu\chi^2_R[f],
\end{equation}
subject to the condition
\begin{equation}
\chi_R^2[f] \le \chi_{\rm exp}^2 = \frac{1}{N}\sum_{i,j}
\sqrt{V(s_i,s_i)V(s_j,s_j)}V^{-1}(s_i,s_j),
\label{mu:cond}
\end{equation}
which will be the criterion to determine the Lagrange multiplier $\mu$.
 This procedure leads to an integral equation for the imaginary part of
 the true amplitude, $f(x;\mu)$:
\begin{equation}
\begin{array}{r}
f(x;\mu)=\displaystyle f_{\rm exp}(x) +
\frac{\lambda|\Gamma_{\rm exp}|}
{\pi|\Gamma_L|}\int_{s_0}^{s_{\rm max}}dy \, V(y,x)
\int_{\Gamma_L}dz \, w_L(z) {\cal C}_n(z,y) {\tilde F}^n_{\rm QCD}(z)
\\
\\
+\displaystyle\lambda\int_{s_0}^{s_{\rm max}}dz \, 
{\cal K}(x,z) f(z;\mu),
\end{array}
\label{inteq}
\end{equation}
where $\lambda=1/\mu$ and
\begin{equation}
{\cal K}(x,z) = -\frac{|\Gamma_{\rm exp}|}{\pi^2|\Gamma_L|}
\int_{s_0}^{s_{\rm max}}dy \, V(y,x)
\int_{\Gamma_L}dx' \, w_L(x') {\cal C}_n(x',y) {\cal C}_n(x',z).
\end{equation}

Thus, the algorithm for determining acceptable values for the
condensates is the following \cite{th:almasy}: 
\begin{itemize}
 \item Choose a model by stating how many terms in the OPE should be
   taken into account. The term with the next-highest dimension is used
   to define an error corridor in the space-like region;
 \item Solve the integral equation (\ref{inteq}) iteratively, with fixed
   values of the free parameters of the chosen model, until the Lagrange
   multiplier $\mu$ satisfies the condition (\ref{mu:cond});
 \item Calculate $\chi_L^2$ corresponding to the above solution as a
   function of the free parameters;
 \item Minimise $\chi_L^2$ with respect to variations of the model
   parameters. The corresponding parameter values are the condensates we
   are looking for;
 \item Determine confidence regions around the fitted parameters by
   solving
   \begin{equation}
    \chi_L^2=\chi_{L,\rm min}^2+\Delta\chi^2.
   \end{equation}
   Here we will assume that the underlying probability distribution is
   Gaussian, fixing $\Delta\chi^2$ to reflect the conventional 1-, 2-
   and $3\sigma$-contours for $n$-parameter fits. Numerical values for
   errors will be given for $1\sigma$ confidence intervals. 
\end{itemize}


\section{Results and discussion}
 
\begin{figure}[b!]
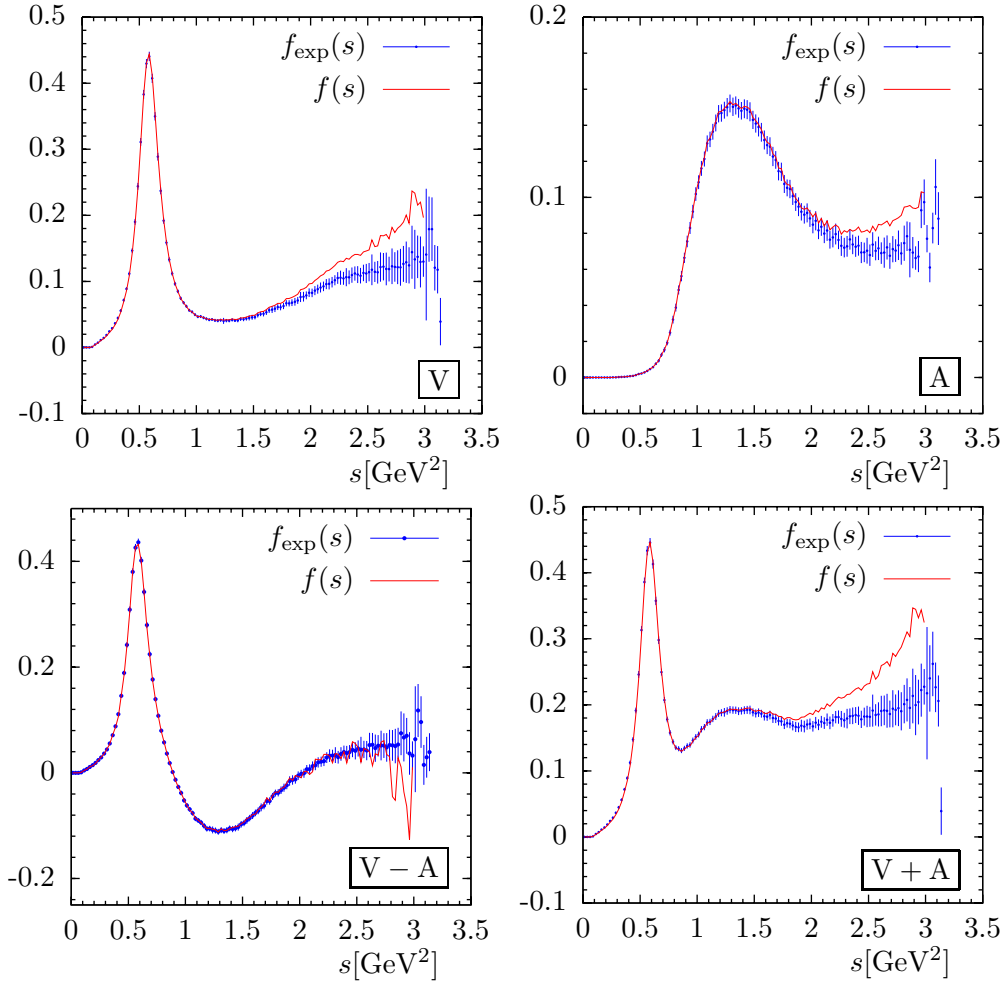

\begin{center}
~\includegraphics{fig1a.epsi} 
~~\includegraphics{fig1b.epsi} 

\includegraphics{fig1c.epsi}  
~~\includegraphics{fig1d.epsi}
\caption{The regularised function $f(s)$ (see Eq.\ (\ref{inteq}))
  compared to the experimental data of Ref.\ \cite{aleph05}. $f(s)$ was
  obtained with 1-parameter fits using the 120 first $s$-bins, i.e.\
  from $s \le 3$ GeV$^2$. }
\label{freg+data}
\end{center}
\end{figure}

The most important step of our analysis is to find a function $f(s)$
 (Eq.\ (\ref{inteq})) which provides a best fit to both the data and the
 theoretical model. A direct comparison of the experimental data with
 the regularised function $f(s)$ obtained from 1-parameter fits is shown
 in Fig.\ \ref{freg+data}. We find nice agreement over the full range of
 $s$ with the exception of the highest $s$-bins. Here the spread of data
 points is apparently wider than individual errors on single data points.
 The largest differences are visible in the $V+A$ channel at $s > 2$ 
 GeV$^2$ where the discrepancies in the $V$ and $A$ channels between data 
 and the fitted function $f(s)$ get enhanced while they appear to be 
 compensated in the $V-A$ channel. We emphasize that we have used the 
 full correlation matrix provided by ALEPH. Correlations are certainly 
 important in our fit, however, they can not be visualized in our figure. 

\subsection{1-parameter fits}

Let us start with 1-parameter fits and quote results for condensates of 
 dimension $d=4$ ($V$, $A$, $V+A$) and $d=6$ ($V-A$) \cite{almasy}: 
\begin{eqnarray}
{\cal O}_4^{V} & = & 1.6_{-0.5}^{+0.4} \times
 10^{-3} {\rm GeV}^4, 
 \hspace*{10mm} \chi_{L,{\rm min}}^2=49, 
\nonumber\\[1ex]
{\cal O}_4^{A} & = & 2.6_{-0.4}^{+0.4} \times 
 10^{-3} {\rm GeV}^4, 
 \hspace*{10mm} \chi_{L,{\rm min}}^2=2.8, 
\nonumber\\[1ex]
{\cal O}_4^{V+A} & = & 4.2_{-0.9}^{+0.8} \times 
 10^{-3} {\rm GeV}^4, 
 \hspace*{10mm} \chi_{L,{\rm min}}^2=19, 
\nonumber\\[1ex]
{\cal O}_6^{V-A} & = & -5.9_{-1.0}^{+1.7} \times 
 10^{-3} {\rm GeV}^6, 
 \hspace*{7mm} \chi_{L,\rm min}^2=0.17.
 \nonumber 
\end{eqnarray}
In the 1-parameter fits we have fixed all higher-dimension condensates
 to be zero.  The results of the $V-A$ analysis had been given earlier
 \cite{almasy}.  There we obtained an acceptable fit when choosing an
 error corridor defined by the dimension $d=8$ condensate in the
 space-like region with
\begin{equation}
 \left|{\cal O}_8^{V-A}\right|_{\rm max} \simeq 1.3\times 
 10^{-3}{\rm GeV^8}.
\label{estO8}
\end{equation}
 Our fit thus provides an indirect estimate of the upper limit of 
 $|{\cal O}_8^{V-A}|$. In the other channels we have used the coefficient
 $K_3$ in the perturbative expansion of the Adler function to define the
 theory error.

\subsection{2-parameter fits}

When performing 2-parameter fits we aim to determine simultaneously the
 first two relevant condensates appearing in the operator product
 expansion. The additional freedom in the fit provided by the second
 parameter allows us, in general, to obtain better fits. In fact, it
 turns out that, except for the $V-A$ channel, the condensates of
 next-to-lowest dimension have significant non-zero values, in contrast
 to the assumption underlying the 1-parameter fits.  

\begin{figure}[t!]
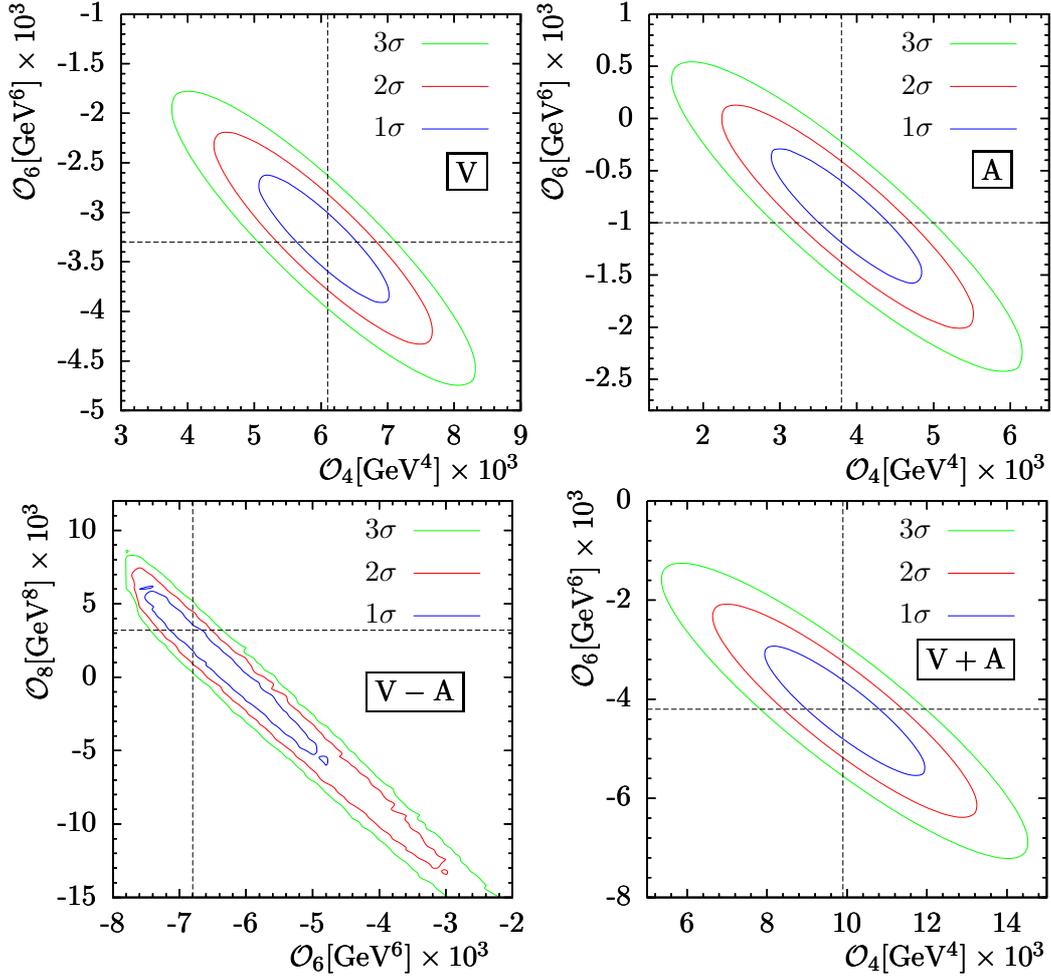

\begin{center}
\includegraphics{fig2a.epsi} 
\includegraphics{fig2b.epsi} 

~\includegraphics{fig2c.epsi} 
\hspace*{3mm}
\includegraphics{fig2d.epsi}
\end{center}
\caption{2-parameter fits: 1-, 2- and 3$\sigma$ confidence regions
  defined by contours of constant $\chi_L^2$ in the plane of the two
  fitted parameters. The central values (see Tab.\
  \ref{twoparamresults}) are marked by dashed lines.}
\label{two:plots}
\end{figure}

In Fig.\ \ref{two:plots} we show contours of constant $\chi^2_L$. One
 can see that we find strong correlations between the two free
 parameters.  This correlation allows us to determine a linear
 combination of the two parameters with a well defined and rather small
 error:
\begin{equation} 
\begin{array}{l}
{\cal O}_6^{V}+0.65\mbox{ GeV}^2\ {\cal O}_4^{V}=
0.66_{-0.25}^{+0.25}\times 10^{-3} \mbox{ GeV}^6,\\
\\
{\cal O}_6^{A}+0.65\mbox{ GeV}^2\ {\cal O}_4^{A}=
1.60_{-0.25}^{+0.26}\times 10^{-3} \mbox{ GeV}^6,\\
\\
{\cal O}_6^{V+A}+0.65\mbox{ GeV}^2\ {\cal O}_4^{V+A}=
2.20_{-0.51}^{+0.50}\times 10^{-3} \mbox{ GeV}^6,
\\ 
\\
{\cal O}_8^{V-A}+2.22\mbox{ GeV}^2\ {\cal O}_6^{V-A}=
-18.30_{-0.25}^{+0.38}\times 10^{-3} \mbox{ GeV}^8.
\end{array}
\end{equation}

The location of the minima, i.e.\ the central values of the fitted
 parameters are quoted in Tab.\ \ref{tab:2param}.  As expected, the
 value of ${\cal O}_8^{V-A}$ found in the 2-parameter fit has the same
 order of magnitude as the corresponding estimate found from the
 1-parameter fit (see Eq.\ \ref{estO8}).  Similarly to the 1-parameter
 fit, we can now give an estimated upper limit of the $V-A$ condensate
 of dimension $d=10$ which was used in the 2-parameter fit to define the
 error channel:
\begin{equation}
 \left|{\cal O}_{10}^{V-A}\right|_{\rm max} \simeq 5.7\times 
 10^{-3}{\rm GeV^{10}}.
\end{equation}
 As before, the error corridors for the $V$, $A$ and $V+A$ channels were
 defined by the perturbative contribution of order $\alpha_s^3$ (Eq.\
\ref{errcorr}). 
 
\begin{table}[t!]
\begin{center}
\begin{tabular}{||c@{\vrule height 14pt depth4pt width0pt
                     \hskip\arraycolsep}|c|c|c|c||}
\hline\hline
 \rule[-2mm]{0mm}{7mm}
 & $V$ & $A$ & $V+A$ & $V-A$ \\ 
 \hline\hline
 \rule[-3mm]{0mm}{8mm}
 $d=4$ & $6.1_{-1.1}^{+0.9}$ & 
         $3.8_{-0.9}^{+1.1}$ & 
         $9.9_{-2.0}^{+2.1}$ & \\ 
 \hline
 \rule[-3mm]{0mm}{8mm}
 $d=6$ & $-3.3_{-0.6}^{+0.7}$ & 
         $-1.0_{-0.7}^{+0.6}$ & 
         $-4.2_{-1.3}^{+1.3}$ & 
         $-6.8_{-0.8}^{+2.0}$ \\ 
 \hline
 \rule[-3mm]{0mm}{8mm}
 $d=8$ & & & & $3.2_{-9.2}^{+2.8}$ \\ 
 \hline
 \rule[-3mm]{0mm}{7mm}
 $\chi_{L,\rm min}^2$ & 20.4 & 0.47 & 7.1 & 0.37\\ 
 \hline\hline
\end{tabular}
\end{center}
\label{twoparamresults}
\caption{2-parameter fits: central values of the fitted parameters in 
 units of $10^{-3}{\rm GeV}^d$ and the corresponding values of  
 $\chi_{L,\rm min}^2$. In the $V$, $A$ and $V+A$ channels the fitted 
 parameters were the condensates of dimension $d=4$ and $6$ while 
 in the $V-A$ channel we have fitted the dimension $d=6$ and $8$ 
 condensates.}
\label{tab:2param}
\end{table}

Despite of the poor agreement of theory with data in the $V$ and $V+A$
 channels, reflected by the large $\chi^2_{L,min}$ values, it is
 important to remark that a consistent over-all picture has emerged from
 our fits.  The results of the 2-parameter fits are in agreement with
 those from the 1-parameter fit.  However, since here the values of
 ${\cal O}^{V,A,V+A}_6$ and ${\cal O}^{V-A}_8$ were left unconstrained,
 we have found larger ranges for ${\cal O}^{V,A,V+A}_4$ and ${\cal
   O}^{V-A}_6$.  Note in particular that the slope of the correlations
 in the $V$, $A$ and $V+A$ channels is the same. For the $V+A$ case, the
 values for both condensates of dimension $d=4$ and $d=6$ agree with the
 values found by actually taking the sum of the results from the $V$ and
 $A$ channels.

\subsection{3-parameter fits}
\label{three:sec}

\begin{figure}[p]
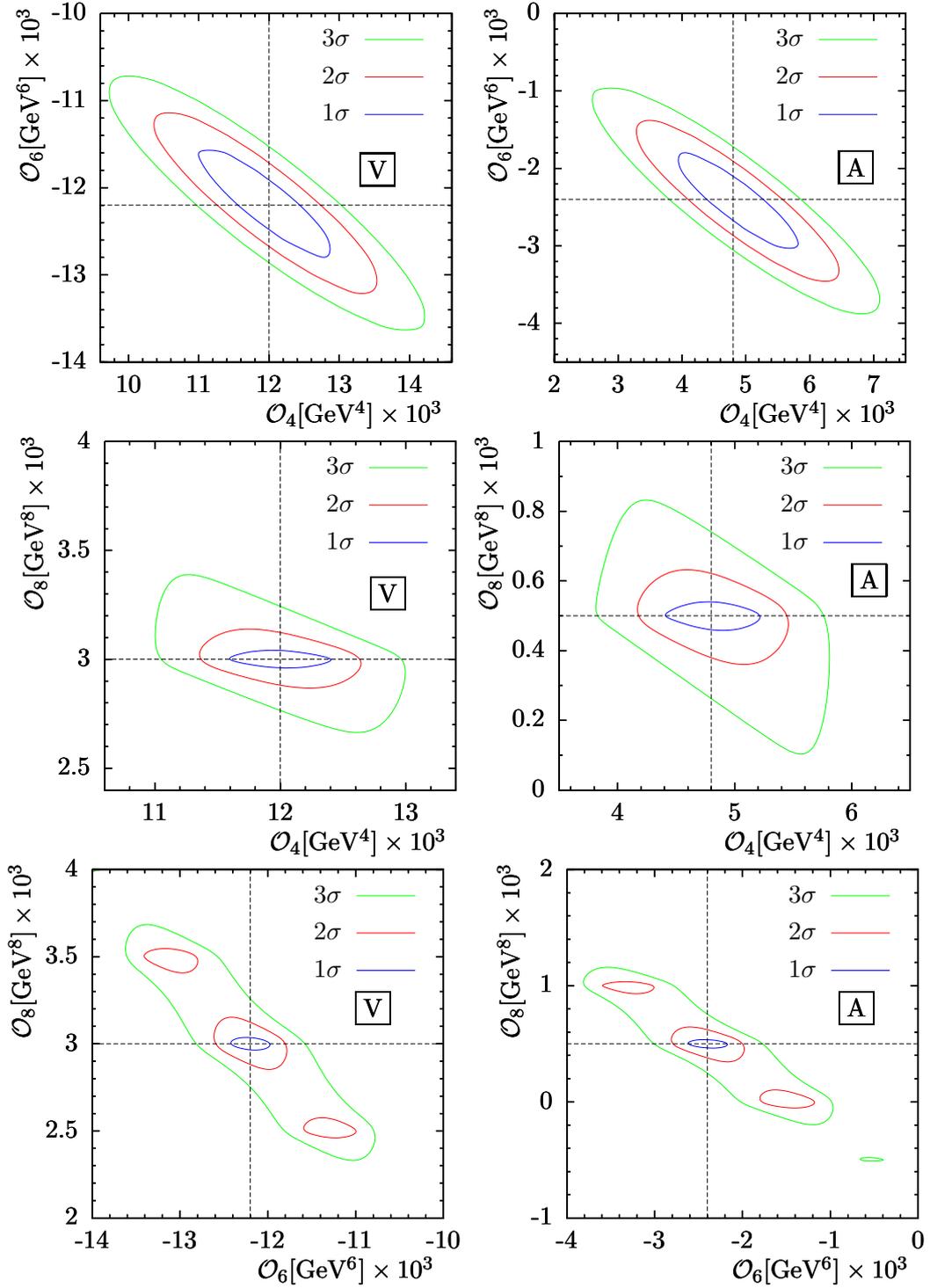

\begin{center}
\includegraphics{fig3a.epsi} 
\hspace*{2mm}
\includegraphics{fig3b.epsi}

~\includegraphics{fig3c.epsi} 
~\includegraphics{fig3d.epsi} 

~\includegraphics{fig3e.epsi} 
\hspace*{3mm}
\includegraphics{fig3f.epsi}
\caption{3-parameter fits in the $V$ and $A$ channels: 1-, 2- and
  3$\sigma$ confidence regions defined by contours of constant
  $\chi_L^2$ in the plane of two of the fitted parameters and located at
  the central value of the third one. The central values (see Tab.\
  \ref{threeparamresults}) are marked by dashed lines.}
\label{three:VA}
\end{center}
\end{figure}

\begin{figure}[p]
\begin{center}
\includegraphics{fig4a.epsi} 
~\includegraphics{fig4b.epsi} 

\includegraphics{fig4c.epsi} 
~\includegraphics{fig4d.epsi}

\includegraphics{fig4e.epsi} 
\includegraphics{fig4f.epsi}
\caption{3-parameter fits in the $V-A$ and $V+A$ channels: 1-, 2- and
  3$\sigma$ confidence regions defined by contours of constant
  $\chi_L^2$ in the plane of two of the fitted parameters and located at
  the central value of the third one. The central values (see Tab.\
  \ref{threeparamresults}) are marked by dashed lines.}
\label{three:VpmA}
\end{center}
\end{figure}

A 3-parameter fit is also possible, where we consider as free parameters
 the first three relevant condensates in the OPE. We have chosen to
 display our results as $\chi_L^2$-contours in the planes defined by the
 three possible pairs of fit parameters. These correlations are shown in
 Fig.\ \ref{three:VA} for the $V$ and $A$ channels and in Fig.\
 \ref{three:VpmA} for the $V\pm A$ channels. In every case we display
 2-dimensional slices of the 3-dimensional allowed parameter ranges
 keeping the third parameter at its central value as obtained from the
 3-parameter fit. These central values are summarised in Tab.\
 \ref{threeparamresults}.

\begin{table}[ht]
\centering
\begin{tabular}{||c@{\vrule height 14pt depth4pt width0pt
                     \hskip\arraycolsep}|c|c|c|c||}\hline\hline
 \rule[-2mm]{0mm}{7mm}
 & $V$ & $A$ & $V+A$ & $V-A$ \\ 
 \hline\hline
 \rule[-3mm]{0mm}{8mm}
 $d=4$  & $12.0_{-1.8}^{+1.6}$ & 
          $4.8_{-1.8}^{+1.8}$  & 
          $16.6_{-3.8}^{+3.2}$ & \\ 
 \hline
 \rule[-3mm]{0mm}{8mm}
 $d=6$  & $-12.2_{-1.8}^{+2.0}$ & 
          $-2.4_{-2.0}^{+2.0}$ & 
          $-14.5_{-4.5}^{+5.0}$ & 
          $-3.2_{-0.4}^{+1.6}$\\ 
 \hline
 \rule[-3mm]{0mm}{8mm}
 $d=8$  & $3.0_{-0.5}^{+0.5}$ & 
          $0.5_{-0.5}^{+0.5}$ & 
          $3.5_{-1.5}^{+1.5}$  & 
          $-17.0_{-9.5}^{+2.5}$\\ 
 \hline
 \rule[-3mm]{0mm}{8mm}
 $d=10$ & & & & $66.0_{-14.0}^{+40.0}$\\ 
 \hline
 \rule[-3mm]{0mm}{7mm}
 $\chi_{L,\rm min}^2$ & 7.15 & 0.17 & 2.51 & 0.35\\ 
 \hline\hline
\end{tabular}
\caption{3-parameter fits: central values of the fitted parameters in 
  units of $10^{-3}{\rm GeV}^d$ and their corresponding values of 
  $\chi_{L,\rm min}^2$. In the $V$, $A$ and $V+A$ channels the fitted 
  parameters were the condensates of dimension $d=4,\ 6$ and $8$, while 
  in the $V-A$ channel we have fitted the dimension $d=6,\ 8$ and $10$ 
  condensates. The error estimates are obtained by projecting the
  3-dimensional ranges allowed by the fit onto the corresponding selected
  parameter. } 
\label{threeparamresults}
\end{table}

An estimated upper limit of the dimension $d=12$ condensate in the $V-A$
 channel, the one used to define the error corridor, is
\begin{equation}
\left|{\cal O}_{12}^{V-A}\right|_{\rm max} \simeq 47\times 
 10^{-3}{\rm GeV^{12}}
\end{equation}
 which is the expected order of magnitude.

Again we observe that 3-parameter fits turn out to provide better
 $\chi^2_{L,min}$ values than the 2-parameter fits. Obviously, the
 improved results are obtained since the higher-dimension condensate can
 be chosen non-zero in the fit. As a consequence, the central values of
 all condensates are shifted. In addition we observe that the
 3-dimensional contours are not always ellipsoids and non-Gaussian
 errors play a role.


\section{Consistency checks}

\begin{figure}[b!]
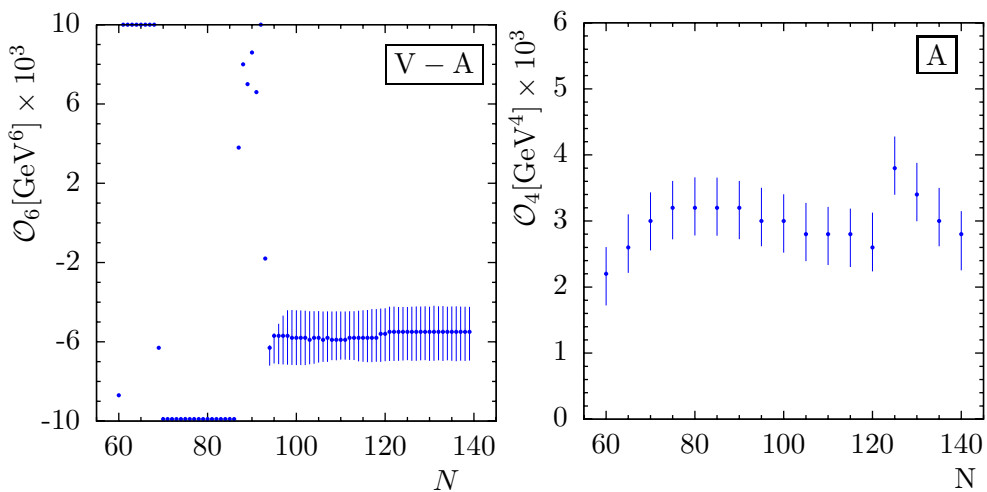

\centering
\includegraphics{fig5a.epsi}
\includegraphics{fig5b.epsi}
\caption{Dependence on the number of data points. $N$ denotes the number
  of the highest data bin used in the analysis.} 
\label{depN}
\end{figure}

In the following we present additional details of our algorithm and 
 describe a number of consistency checks. We have, in particular, studied 
 the behaviour of the algorithm and its results with respect to 
 variations of parameters appearing in the analysis: the number of 
 experimental data points $N$ used for the extraction of condensates, the 
 end-points of the time-like interval $\Gamma_L$, $s_1$ and $s_2$, as well 
 as the dependence on the error parameter needed to define $\sigma_L$. 
 In this section, we restrict ourselves to the case of 1-parameter fits.

\begin{figure}[t!]
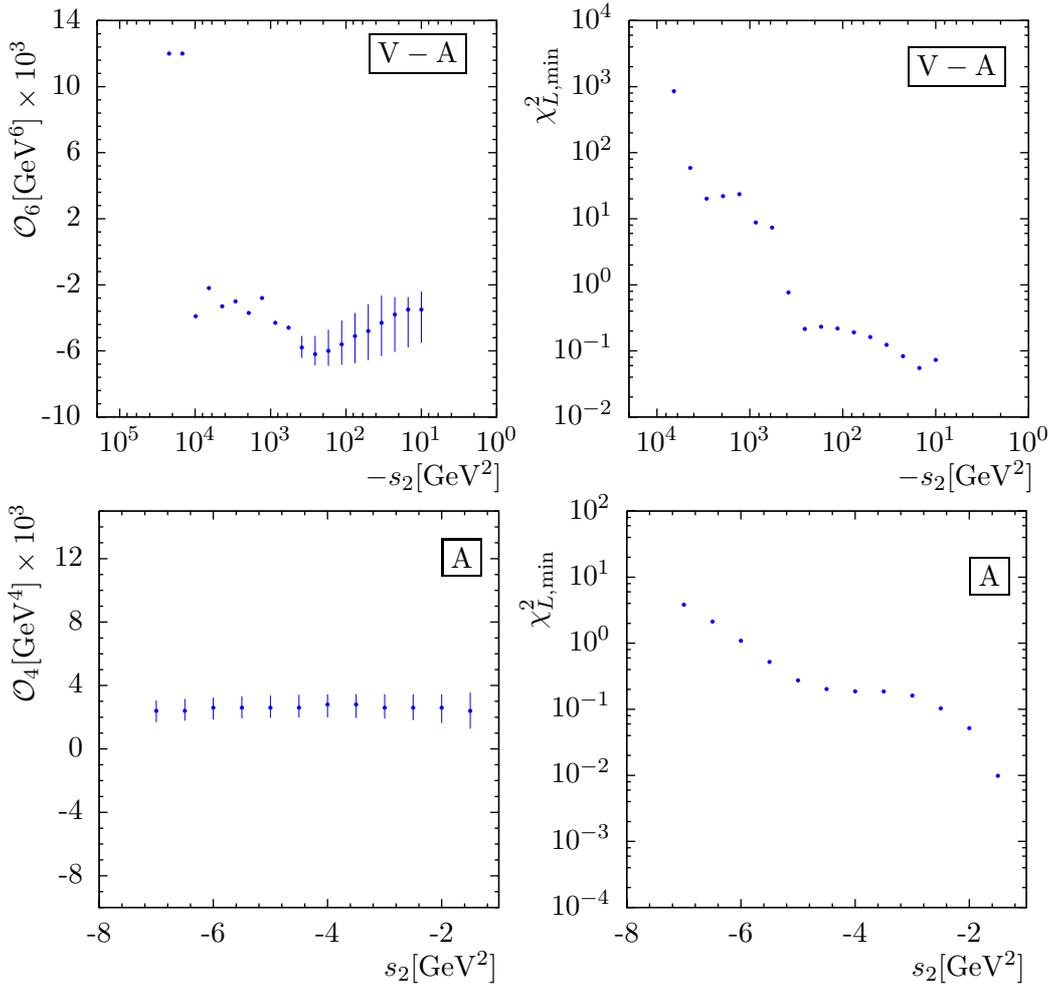

\begin{centering}
~~\includegraphics{fig6a.epsi}  
\includegraphics{fig6b.epsi} 

\includegraphics{fig6c.epsi} 
~\includegraphics{fig6d.epsi} 

\caption{Dependence on the lower end, $s_2$, of the space-like interval:
 the fitted parameter as a function of $s_2$ (left) and $\chi_{L,{\rm
     min}}^2$ as a function of $s_2$ (right). 
}
\label{deps2}
\end{centering}
\end{figure}

One can show that the information on the condensates is contained in the
 lower part of the spectrum by adding or removing data points at largest 
 $s$. In Fig.\ \ref{depN} (left panel) we show how the fit result for
 ${\cal O}_6^{V-A}$ depends on the number of data points. One can
 observe a fast stabilisation of the result already for the $N \simeq
 100$ lowest-$s$ data points. In contrast, for the case of the
 $A$-correlator, cf.\ Fig.\ \ref{depN} (right panel), one can observe
 that including or excluding data points above $N = 120$ has a stronger
 effect on the condensate ${\cal O}_4^{A}$. In this region the rapid
 oscillation of the data points as well as large experimental errors
 play an important role. The decision not to include experimental
 results from the highest bins in the analysis, is supported by
 inspecting Fig.\ \ref{freg+data}: there we found that the regularised
 function obtained in our analysis does not describe the data in the
 large-$s$ region. We have thus decided to cut off data points above $s
 = 3$ GeV$^2$, i.e., we use only the $N=120$ first data points.

Figure \ref{deps2} shows the behaviour of the algorithm with respect to
 changes of $s_2$, the lower limit of the space-like interval $\Gamma_L$. 
 One should choose $|s_2|$ as large as possible, but we observe stability 
 of the algorithm for values certainly not larger than a few times $10^2$ 
 GeV$^2$ for the $V-A$ analysis and even smaller values are required for 
 the fits of the $A$ condensate. This is illustrated in the right column 
 of Fig.\ \ref{deps2} which shows the dependence of $\chi^2_{L,\rm min}$ 
 on $s_2$. One can observe a plateau for the values of ${\cal O}_6^{V-A}$ 
 and ${\cal O}_4^{A}$ as a function of $s_2$ and thus infer the values 
 used in the analysis to be $s_2 = -150$ GeV$^2$ for $V-A$ and $s_2 = - 
 3.5$ GeV$^2$ for the $A$ channel. For larger values of $|s_2|$ the 
 minimum of $\chi_L^2$ becomes larger than 1, signaling a bad fit. This
 behaviour may be due to numerical instabilities and limitations of 
 experimental data; more important, however, is the fact that in our LO
 analysis we did not take into account perturbative higher-order
 corrections: the large perturbative tails of the $V$- and
 $A$-correlators become increasingly important when increasing $|s_2|$
 and the sensitivity to the low-energy condensates is lost.

\begin{figure}[b!]
\centering
\includegraphics{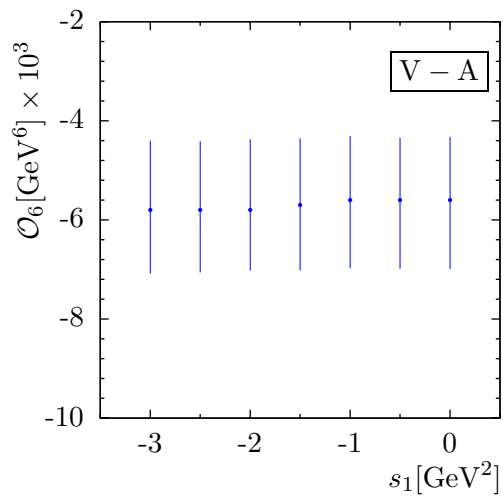}
\caption{Dependence on $s_1$, the upper limit of the space-like interval
 $\Gamma_L$ for ${\cal O}_6^{V-A}$.}
\label{deps1}
\end{figure}

In the case of the $A$-correlator, when studying the dependence on
 $s_2$, we found that the best simultaneous description of experimental
 data and theory is obtained when we choose to define the error corridor 
 with the help of the last known term in the perturbation series, i.e., 
\begin{equation}
 \sigma_L^{A}(x)=
 \displaystyle\frac{1}{4\pi^2}K_3\left(
\frac{\alpha_s(-x)}{\pi}\right)^3.
\label{aerrorchannel}
\end{equation}
 In contrast, for an error corridor calculated from the next-higher term
 in the OPE, i.e., using $\sigma_L^A(s) = \mbox{const} / s^d$, we
 observe a less distinct plateau when varying $s_2$ and no stability for
 the results for ${\cal O}_4^{A}$. The fact that with the choice
 (\ref{aerrorchannel}) we find very stable fit results for ${\cal
   O}_4^{A}$ even when increasing $|s_2|$ beyond the range of the
 $\chi^2_{L,min}$-plateau, makes us confident that our results for the
 $A$-correlator are meaningful.

It is, in fact, to be expected that a definition of the error corridor
 with the help of the higher-dimensional terms in the OPE would narrow
 too fast (with a power of $1/s$) and not leave space enough for
 perturbative contributions that fall only logarithmically. In contrast,
 for the $V-A$ channel where perturbative contributions are absent, it
 was possible to choose tighter error channels given by the omitted
 next-higher OPE term. We note that the definition of the error channels
 was the same for all 1-, 2-, and 3-parameter fits in the case of the
 $V$, $A$ and $V+A$ channels, whereas for the $V-A$ channel we had to
 re-define the error channel when increasing the number of free
 parameters. This explains the observation that $\chi^2_{L,min}$ is not
 necessarily increasing when going from 1- to 2- and to 3-parameter fits
 for $V-A$ condensates.

There exists also a well defined plateau for the fitted parameters as a
 function of $s_1$. In the analysis we have chosen to use the values 
 $s_1 = -1.0$ GeV$^2$ for $V-A$ and $s_1 = -0.4$ GeV$^2$ for the
 $A$ channel (see Fig.\ \ref{deps1}).

All these consistency checks were performed for the $V$ and $V+A$ channels 
 as well with similar results and we found no justification to change the 
 number of data points $N$ used in the analysis or $s_1$, the upper limit 
 of the space-like interval $\Gamma_L$. Also, the dependence on the lower 
 limit $s_2$ has shown that the best simultaneous description of theory 
 and data corresponds to an error corridor defined by the last known term 
 in the perturbative series. For the correlator of the vector current,
 however, we are not convinced that we have obtained trustworthy
 results: first, $\chi^2_{L,min}$ is large even for a 3-parameter fit
 and, second, the fit results for ${\cal O}_4^{V}$ and ${\cal O}_6^{V}$
 change by more than the estimated uncertainties when changing the
 number of free parameters in the fit. Therefore we do not discuss
 results for the $V$-correlator further. Note, however, that ${\cal
   O}_4^{V}$ and ${\cal O}_4^{A}$ are predicted to be equal, and fit
 results for one can be used to determine the other.


\section{Comparison with other results and conclusions}

There exists a number of previous extractions of QCD condensates in the
 literature, mainly based on sum rule approaches. For the $V-A$ channel 
 they are listed in Tab.\ \ref{LOdet} together with a repeated collection 
 of our results.

\begin{table}[b!]
\begin{center}
\begin{tabular}{||l@{\vrule height 14pt depth4pt %
                            width0pt\hskip\arraycolsep}|c|c|c|c||c}
\hline\hline
\rule[-3mm]{0mm}{8mm}
   & 
   ${\cal O}_6^{V-A}$ & 
   ${\cal O}_8^{V-A}$ & 
   ${\cal O}_{10}^{V-A}$ & 
   ${\cal O}_{12}^{V-A}$ \\ 
\hline
\rule[-2mm]{0mm}{6mm}
\cite{rojo} & 
   $-4\pm2.0$ & 
   $-12_{-11}^{+7}$ & 
   $78\pm24$ & 
   $-2.6\pm0.8$\\ 
\hline 
\rule[-2mm]{0mm}{6mm}
\cite{bordes}$^\ast$ & 
   $-4.52\pm1.1$ & 
   $-10.8\pm6.6$ & 
   $72\pm28$ & 
   $-240\pm100$\\ 
\hline 
\rule[-2mm]{0mm}{6mm}
\cite{cirigliano1}$^\ast$ & 
   $-2.27\pm0.51$ & 
   $-2.85\pm2.18$ & 
   $24.1\pm6.1$ & 
   $-80\pm16$\\ 
\hline 
\rule[-2mm]{0mm}{6mm}
\cite{narison1} & 
   $-8.7\pm2.3$ & 
   $15.6\pm4.0$ & 
   $-17.1\pm4.4$ & 
   $14.7\pm3.7$\\ 
\hline 
\rule[-2mm]{0mm}{6mm}
\cite{friot} & 
   $-7.9\pm1.6$ & 
   $11.7\pm2.6$ & 
   $-13.1\pm3.0$ & 
   $13.2\pm3.3$\\ 
\hline 
\rule[-2mm]{0mm}{6mm}
\cite{zyablyuk} & 
   $-7.2\pm1.2$ & 
   $7.8\pm2.5$ & 
   $-4.4\pm2.8$ & 
   \\ 
\hline 
\rule[-2mm]{0mm}{6mm}
\cite{dominguez}$^\ast$ & 
   $-8\pm2$ & 
   $-2\pm12$ & 
   & 
   \\ 
\hline 
\rule[-2mm]{0mm}{6mm}
\cite{ioffe} & 
   $-6.8\pm2.1$ & 
   $7\pm4$ &  
   & 
   \\ 
\hline 
\rule[-2mm]{0mm}{6mm}
\cite{aleph} & 
   $-7.7\pm0.8$ & 
   $11.0\pm1.0$ &  
   & 
   \\ 
\hline 
\rule[-2mm]{0mm}{6mm}
\cite{opal} & 
   $-6\pm0.6$ & 
   $7.5\pm1.3$ &  
   & 
   \\ 
\hline 
\multicolumn{5}{||l||}{\rule[-2mm]{0mm}{7mm} This work}\\
\hline
\rule[-2mm]{0mm}{6mm}
1-parameter fit \cite{almasy}& 
   $-5.9_{-1.0}^{+1.7}$ &  
   &  
   & 
   \\ 
\hline 
\rule[-2mm]{0mm}{6mm}
2-parameter fit \cite{almasy}& 
   $-6.8_{-0.8}^{+2.0}$ & 
   $3.2_{-9.2}^{+2.8}$ &  
   & 
   \\ 
\hline 
\rule[-2.5mm]{0mm}{6.5mm}
3-parameter fit & 
   $-3.2_{-0.4}^{+1.6}$ & 
   $-17.0_{-9.5}^{+2.5}$ & 
   $66.0_{-14.0}^{+40.0}$ & 
   \\ 
\hline \hline
\end{tabular}
\caption{Estimated values of the condensates ${\cal O}_d^{V-A}$ of 
 dimension $d\le12$ in units of $10^{-3}$ GeV$^d$ at leading order.
 References marked with a {}$^\ast$ use a different normalisation of 
 spectral functions. The values shown are adjusted so that they can be 
 compared to those of the present work.}
\label{LOdet}
\end{center}
\end{table}

In most cases, errors given by the authors are in the order of $25\%$, 
 sometimes even as small as $10\%$. For the $d=6$ condensate, our results 
 fall nicely in the same range, also with an error estimate which is 
 comparable to that of other analyses. The spread of the central values 
 is, however, larger than the typical error. We believe that the observed 
 variation of these results represent the ambiguities inherent in the 
 QCD sum rule approach. 

The situation is more difficult to summarize in the case of the 
 higher-dimensional $V-A$-condensates: the variation of results from
 different analyses is even bigger, but estimates of relative errors are
 again in some cases similar to those of the $d=6$ condensates. A
 possible reason for this inconclusive picture may be related to the
 strong correlation between condensates of different dimension.
 Consider, for example, our results for ${\cal O}_8^{V-A}$.  The 2- and
 3-parameter fits lead to very different values since the assumptions
 underlying the two fits are different: in the 2-parameter fit we
 assumed ${\cal O}_{10}^{V-A} = 0$, whereas the 3-parameter fit
 preferred the value ${\cal O}_{10}^{V-A} = 66$ GeV$^{10}$ and the range
 of values for ${\cal O}_8^{V-A}$ given in the table is for that fixed
 central value of ${\cal O}_{10}^{V-A}$.

It is also interesting to note the agreement of the correlation between
 ${\cal O}_6^{V-A}$ and ${\cal O}_8^{V-A}$ found in our analysis with
 corresponding results from \cite{narison1,zyablyuk}. In Ref.\
 \cite{narison1}, the linear combination of these two parameters is
 extracted from weighted finite energy sum rules, but no errors were
 given, while in Ref.\ \cite{zyablyuk} Borel sum rules were used to find
 the correlation. In the latter reference, 1-, 2- and 3$\sigma$
 confidence regions for the correlations ${\cal O}_6^{V-A}$--${\cal
   O}_8^{V-A}$ and ${\cal O}_6^{V-A}$--${\cal O}_{10}^{V-A}$ are
 presented. The $1\sigma$ contours for ${\cal O}_6^{V-A}$ and ${\cal
   O}_8^{V-A}$ are shifted as compared to ours, but the slope agrees
 well within errors. A careful analysis shows that there is also
 agreement for the ${\cal O}_6^{V-A}$--${\cal O}_{10}^{V-A}$ correlation
 with the result of Ref.\ \cite{zyablyuk}. There, a positive correlation
 was found from a 2-parameter fit which corresponds to fix ${\cal
   O}_8^{V-A} = 0$. With the same assumption we find a correlation of
 the same sign, however a smaller slope. Note that the correlation as
 shown in Fig.\ \ref{three:VpmA} (left column) appears to be different
 when fixing ${\cal O}_8^{V-A}$ at its central value which was found to
 be 3.2, i.e.\ significantly different from zero, in our 3-parameter fit.  

There also exist some previous extractions of QCD condensates in the
 $V$ and $A$ channels, again based on sum rule approaches. The
 normalisation of spectral functions is different from ours and there is
 also a factor of $8\pi^2$ absorbed in the definition of the condensates. 
 We have translated the results so that they can be compared to ours and 
 summarised them for the axial-vector correlator in Tab.\ \ref{VAdet}.

One can remark that the majority of the values found in this work are
 consistent with those from the literature. The sign of ${\cal O}_6^A$,
 though, disagrees with the vacuum saturation approximation and with the
 results from \cite{dominguez0}.
 
As a conclusion, we can state that the values and ranges found for the QCD
 condensates are consistent among themselves and, partly, with previous
 extractions found in the literature even though the agreement between
 theory and data is very poor in the case of the $V$ and $V+A$ channels. 
 Since at present the analyses are still subject to a number of 
 restrictions, one can hope that future work will allow us to improve the 
 agreement between theory and data further. 

\begin{table}[t!]
\begin{center}
\begin{tabular}{||l@{\vrule height 14pt depth4pt %
                            width0pt\hskip\arraycolsep}|c|c|c||c}
\hline\hline
\multicolumn{4}{|c||}{$A$ channel \rule[-2mm]{0mm}{7mm}}\\ 
\hline
\rule[-2mm]{0mm}{6mm}
   & 
   ${\cal O}_4^{A}$ & 
   ${\cal O}_6^{A}$ & 
   ${\cal O}_8^{A}$\\ 
\hline
\rule[-2mm]{0mm}{6mm}
\cite{dominguez0} & 
   $(1.2 \ldots 2.5)$ & 
   $(4.1 \ldots 7.1)$ & 
   $- (19.1 \ldots 13.7)$ \\ 
\hline 
\rule[-2mm]{0mm}{6mm}
\cite{dominguez1} & 
   $(1.3 \ldots 4.8)$ & 
   $- (16.5 \ldots 1.3)$ & 
   \\ 
\hline 
\multicolumn{4}{||l||}{This work \rule[-2mm]{0mm}{7mm}}\\
\hline
\rule[-3mm]{0mm}{8mm}
1-parameter fit & 
   $2.6_{-0.4}^{+0.4}$ & 
   &  
   \\ 
\hline 
\rule[-3mm]{0mm}{8mm}
2-parameter fit & 
   $3.8_{-0.9}^{+1.1}$ & 
   $-1.0_{-0.7}^{+0.6}$ &  
   \\ 
\hline 
\rule[-3mm]{0mm}{8mm}
3-parameter fit & 
   $4.8_{-1.8}^{+1.8}$ & 
   $-2.4_{-2.0}^{+2.0}$ & 
   $0.5_{-0.5}^{+0.5}$\\ 
\hline\hline 
\end{tabular}
\caption{Estimated ranges for the dimension $d\le8$ condensates of the
  $A$ channels in units of $10^{-3}$ GeV$^d$ at leading order.  Existing
  results from the literature are presented. Note that the
  normalisation, for all of them, was adjusted so that they can be
  compared to those from this work.}
\label{VAdet}
\end{center}
\end{table}

When analysing all four channels, we have assumed chiral symmetry,
 decoupling of heavy quarks and the absence of duality violations. If the 
 chiral symmetry is broken, there are also lower-order terms entering the 
 OPE and also mass terms would be present both in the OPE and the 
 perturbative expansion. Moreover, there will be also a perturbative 
 contribution to the $V-A$-correlator. Also the inclusion of heavy quarks 
 is expected to play an important role at high energies. Their 
 contribution would modify the evaluation of the theory prediction 
 ${\tilde F}^n_{\rm QCD}(s)$ in Eq.\ (\ref{dispsource}). It remains to be 
 seen whether these effects are negligible or not. 

Duality refers to the assumption that the true function $\Pi(s)$ can be 
 replaced without error by the expression given by the operator product 
 expansion, $\Pi_{\rm OPE}(s)$. The term {\em duality violation} refers 
 to any contribution missed by the substitution $\Pi(s) \rightarrow 
 \Pi_{\rm OPE}(s)$. As stated already, in our analysis we have assumed 
 that duality violations are absent. It is an interesting task to check 
 how the results would change if one would consider duality violating  
 contributions. Unfortunately, little is known about the structure of 
 duality violations in QCD and one has to rely on model assumptions like 
 those of Ref.\ \cite{cata}. At the time being, possible deviations from 
 duality are suspected to be a major source of theoretical uncertainties 
 \cite{shifman4,zakharov}.     

\section*{Acknowledgements}

A.\ A.\ Almasy would like to thank the Graduiertenkolleg {\em 
 ''Eichtheorien -- Experimentelle Tests und theoretische Grundlagen''} 
 for financial support during the time this work was done.


\end{document}